\documentclass[%
 reprint,
showpacs,
 amsmath,amssymb,
 aps,
]{revtex4-1}

\usepackage{graphicx}% Include figure files
\usepackage{dcolumn}% Align table columns on decimal point
\usepackage{bm}% bold math
\usepackage{textcomp}
\usepackage{bm,latexsym,amsfonts,amsmath,amssymb,amsthm,url,graphics,psfrag,amscd,stmaryrd, mathrsfs}
\usepackage[english]{babel}
\usepackage[textsize=small]{todonotes}       % includes TO DO LIST
\usepackage{textcomp}
\usepackage{mathrsfs} 
\usepackage{todonotes}
\usepackage{color}
\usepackage{graphics}
\usepackage{graphicx}

\usepackage{color}

\newcommand {\be} {\begin{equation}}
\newcommand {\ee} {\end{equation}}

\newcommand {\bea}  {\begin{eqnarray}}
\newcommand {\eea}  {\end{eqnarray}}

\newcommand {\bdm} {\begin{displaymath}}
\newcommand {\edm} {\end{displaymath}}

\newcommand{\p}{{\boldsymbol{p}}}
\newcommand{\q}{{\boldsymbol{q}}}

\newcommand{\pr}{\mathbb P}

\newcommand{\N}{\mathbb N}
\newcommand{\stq}{\left [ s^\star\right ]_{q^\star}}

\newcommand{\e}{{\rm e}}

\newcommand{\atanh}{{\rm atanh\,}}
\newcommand{\eqn}[1]{\begin{equation}#1\end{equation}}
\newcommand{\eqan}[1]{\begin{align}#1\end{align}}

\newcommand{\ch}{\textrm{cosh}}

\newcommand{\kL}{\mathcal{L}}

\newcommand{\bla}{\big \langle}
\newcommand{\bra}{\big \rangle}
\newcommand{\lla}{\left \langle}
\newcommand{\rra}{\right \rangle}

\newcommand{\asinh}{{\rm asinh}}
\newcommand{\nn}{\nonumber}
\newcommand{\chr}[1]{\textcolor{black}{#1}\textcolor{black}}

\begin{document}

\preprint{APS/123-QED}

\title{Annealed inhomogeneities in random ferromagnets}
%\title{ The annealed Ising model on complex networks}
\author{Van Hao Can }
%\email{cvhao89@gmail.com}
\affiliation{ 
Institute of Mathematics, Vietnam Academy of Science and Technology, 18 Hoang Quoc Viet, 10307 Hanoi, Vietnam.}
\author{Cristian Giardin\`a}
%\email{cristian.giardina@unimore.it}
\affiliation{University of Modena and Reggio Emilia, via Universit\`a 4, 41121 Modena, Italy}
%G. Campi 213/b, 41125 Modena, Italy}
%\affiliation{Dipartimento di Fisica, Informatica e Matematica,  University of Modena and Reggio Emilia, via G. Campi 213/b, 41125 Modena, Italy}
\author{Claudio Giberti}
\affiliation{University of Modena and Reggio Emilia, via Universit\`a 4, 41121 Modena, Italy}
%\email{claudio.giberti@unimore.it}
%\affiliation{
%Dipartimento di Scienze e Metodi dell'Ingegneria,
%Universit\`a di Modena e Reggio Emilia,  Via Amendola 2, Padiglione Morselli, I-42122 Reggio E., Italy.}
\author{Remco van der Hofstad}
%\email{r.w.v.d.hofstad@tue.nl}
\affiliation{Eindhoven University of Technology, P.O. Box 513, 5600 MB Eindhoven, The Netherlands}

%\date{\today}

\begin{abstract}
\centerline{\bf Abstract}
We consider spin models on complex networks frequently used to model 
social and technological systems. We study the {\em annealed} ferromagnetic Ising model for random 
networks with either independent edges (Erd\H{o}s-R\'enyi), or with prescribed degree distributions 
(configuration model). Contrary to many physical models, the annealed setting is poorly understood 
and behaves quite differently than the quenched system. In annealed networks with a fluctuating 
number of edges, the Ising model changes the degree distribution, an aspect previously ignored. 
For random networks with Poissonian degrees,
this gives rise to {\em three distinct} annealed critical temperatures depending on the precise model choice, 
only one of which reproduces the quenched one. 
In particular, two of these annealed critical temperatures are {\em finite} even when the quenched one 
is infinite, since then the annealed graph creates a giant component for all sufficiently small temperatures.
We see that the critical exponents in the configuration model with {\em deterministic degrees} are  
the same as the quenched ones, which are the mean-field exponents if the degree distribution has finite fourth moment, 
and power-law-dependent critical exponents otherwise. Remarkably, the annealing for the configuration model with 
{\em random i.i.d.\ degrees} washes away the universality class with power-law critical exponents.
\end{abstract}

%\pacs{05.40.-a, 45.50.-j, 02.50.Ey, 05.70.Ln, 05.45.-a}
%\ams{82C22, 82C35, 34C14}

\maketitle

In spin systems with disorder usually two averaging procedure 
are considered: the {\em quenched} state, that is used to model 
the setting where the couplings between spins are essentially frozen
and the {\em annealed} state, in which  spins and disorder are treated
on the same footing. In this paper we ask the following question:
how different are the {quenched} and {annealed} states of
a {\em disordered ferromagnet}? Do they share the same critical
temperatures and critical exponents?
We show here that this seemingly simple question
does not admit a simple answer. Instead, the comparison
of the annealed  state of a random ferromagnet to the quenched one 
reveals a host of surprises. As we shall see by considering several models
of random graphs, the answer depends sensitively on whether the {\em total number
of edges} of the underlying random graph is fixed, or is allowed
to fluctuate.  Indeed, the typical graph under the annealed measure re-arranges itself in order to 
maximize the ferromagnetic alignment of spins 
by increasing the number of edges.
As a consequence, we argue that the {\em annealed 
critical temperature} is highly model-dependent, even 
in the case of graphs that are asymptotically equivalent (such as the
different versions of the simple Erd\H{o}s-R\'enyi random graph). 
This is to be contrasted
to the {\em quenched critical temperature} that is essentially 
the same for all locally-tree like graphs.

The difference between quenched and annealed becomes
even more substantial in the presence of inhomogeneities that
produce a fat-tail degree distribution,
whose tail behavior is characterized by a 
power-law exponent $\tau >2$.
In this case it has been shown \cite{leone,doro-zero} that quenched models, 
on top of the mean-field universality class, may have
other university classes, where the
quenched critical exponents depend on the power-law exponent
$\tau$, taking the mean-field values for  $\tau >5$, but different
values for $\tau\in(3,5)$.

Our analysis shows that the picture radically changes 
in the annealed setting. In the context of the 
{\em configuration model} we find that 
when the degrees are fixed, one obtains the
same universality classes as in the quenched setting.
However the annealed partition function of the configuration model 
with random (i.i.d.) degrees blows up for fat-tail degree
distributions. Furthermore, for models having
a well-defined partition function, 
\begin{center}
{\em the power-law
universality classes are washed away,}
\end{center} 
and only the mean-field universality class survives.   
%All in all, the increase of randomness 
%in the annealed graph leaves only the mean-field universality class.

The distinction between quenched and annealed averaging 
is particularly relevant for social systems, where the network of acquaintances of members
of a group changes quite rapidly,  on a time scale that is comparable to that of opinion changes
\cite{bara, nbw,decelle}.
In such settings, the annealed setting is the most appropriate.

The multi-facetted phenomenology that we find in the description of the annealed state 
did not emerge in previous studies of disordered
ferromagnets \cite{bianconi,doro-zero,GIU1, lee, leone}, that instead suggested  the annealed state
to be described by an approximate mean-field  theory  
that accounts for heterogeneity of the graph
(so-called``{\em annealed network approach}'' \cite{doro-zero}).
Our analysis shows that this approximate theory may fail
to  identify the critical temperature, 
even for very simple random graph models. 
In the following we first discuss the case of 
homogeneous models with Poissonian degrees 
and, afterwards, we extend our analysis to 
models with inhomogeneities. 

\medskip
\noindent
{\bf Models with Poisson degree distributions.}
Let us consider the Ising model on a network with $n$ vertices. 
Given a spin configuration $\sigma=(\sigma_1, \ldots , \sigma_n) \in \{-1,+1\}^n$ 
and a random graph  with vertex set $V$ and edge set $E$, the Hamiltonian is defined as
\be\label{ham11}
H_n(\sigma)=\beta  \hspace{-0.1cm}\sum_{(i,j)\in E} \sigma_i \sigma_j + B \sum_{i\in V} \sigma_i
\ee
where $\beta$ is the inverse temperature and $B$ an external field.
The order parameter is the spontaneous {\em annealed magnetization}
$M(\beta) = \lim_{B\to 0^+} M(\beta,B),$ where
\begin{equation}
\label{magn-an}
M(\beta,B) = \lim_{n\to\infty} \frac
{ \Big \langle \sum_{\sigma} \Big( \frac{1}{n}\sum_{i=1}^n \sigma_i \Big ) \e^{H_n(\sigma)}\Big \rangle}
{\Big \langle \sum_{\sigma}\e^{H_n(\sigma)}\Big \rangle}\;.
\end{equation}
Here $\langle \cdot \rangle$ denotes expectation over the randomness of the graph,
which, \chr{in the annealed setting,}{} appears both in \chr{numerator and denominator.}{}

The simplest possible random network is the {\em binomial  Erd\H{o}s-R\'enyi}, denoted   $\text{bER}(\lambda/n)$, in which a pair of vertices 
\chr{in $[n]=\{1, \ldots, n\}$}{} is connected (independently from other pairs) with the same probability $\lambda/n,\, \lambda>0$. 
In this case, the Hamiltonian \eqref{ham11} becomes
\be\label{ham}
H^{\text{bER}(\lambda/n)}(\sigma)=\beta  \hspace{-0.3cm}\sum_{1\le i < j \le n} I_{i,j} \sigma_i \sigma_j + B \sum_{i=1}^n \sigma_i,
\ee
where $I_{i,j}$ are  independent and identically distributed Bernoulli random variables with $\pr(I_{i,j}=1)=\lambda/n$
defining the adjacency matrix of the network.
The random variable $D_i =  \sum_{k} I_{i,k}$, 
counting the number of edges connected to vertex $i$ is the {\em degree}  of $i$,
which for large $n$ results in a Poisson random variable with parameter $\lambda$.
As shown in the Supplementary Material, the annealed magnetization 
\eqref{magn-an}  of the  {\em binomial  Erd\H{o}s-R\'enyi}  model
solves the mean-field Curie-Weiss equation
with a renormalized temperature
$\beta\mapsto\sinh(\beta)$, i.e.,
\be
\label{magn-er-bin}
M^{\text{bER}} = \tanh\Big(\lambda \sinh(\beta) M^{\text{bER}}  + B\Big),
\ee
yielding a critical inverse temperature
\be
\beta_c^{\text{bER}}= \asinh(1/\lambda),
\ee
and critical exponents of the mean-field universality class.

The ``{annealed network approach}''  introduced in \cite{bianconi, doro-zero} is based 
on the idea of replacing the model with Hamiltonian \eqref{ham11} by a {mean-field model
on a weighted fully connected graph} described by the Hamiltonian
\be\label{H_Dor-xxx}
H^{\mathrm{mf}}(\sigma) = \frac{\beta}{2} \sum_{i,j=1}^n \frac{D_i D_j}{\ell_n} \sigma_i\sigma_j + B  \sum_{i=1}^n \sigma_i,
\ee
where $\ell_n=\sum_{i}D_i$. 
This translates into an equation for the magnetization given by
\be
\label{magn-dor}
M^{\mathrm{mf}}(\beta,B) = \lla \tanh\Big(\beta y D + B\Big)\rra,
\ee
where
$y \in (0,1)$  is a solution of the mean-field equation
\be
\label{dor-fixed}
y = \lla{D}\tanh\Big(\beta y D + B\Big)\rra /\bla D\bra\,.
\ee
In the case of models with Poissonian degrees with $\bla D\bra = \lambda$,
the linearization around $y=0$ yields the inverse
critical temperature
\be
\beta_c^{\mathrm{mf}}= 1/(\lambda+1),
\ee
and critical exponents are those of the mean-field 
universality class. Therefore, the ``annealed network approach'' 
predicts the correct annealed critical exponents, \chr{but}{} 
fails in determining the critical temperature. As shown in Fig.\ 1, the discrepancy
between the true value of the critical temperature
(red curve) and the one predicted by 
the annealed network approach (blue curve)
increases as the average connectivity 
$\lambda$ is decreased. In particular
at $\lambda =0$ one gets 
$\beta_c^{\mathrm{mf}} = 1$, \chr{which}{}
is clearly unphysical.

It is interesting to compare the 
annealed magnetization \eqref{magn-an} to the
magnetization that is obtained 
in the {\em quenched setting} \cite{GGHP2} .
For all the models that are locally tree-like  
 \cite{leone,doro-zero,dm,DGGHP},
the quenched magnetization $M^{\mathrm{qu}}(\beta,B)$ 
is 
\begin{equation}
\label{m-quenched}
M^{\mathrm{qu}} (\beta, B) = \lla \frac{\e^{2B}-\prod_{i=1}^D X_i}{\e^{2B}+\prod_{i=1}^D X_i}\rra,
\end{equation}
where $(X_i)_{i\geq 0}$ are i.i.d.\ random variables satisfying
\be
X_0 \overset{(\kL)}{=} \frac{\e^{-\beta +B} +\e^{\beta-B} \prod_{i=1}^{D} X_i }{\e^{\beta +B} +\e^{-\beta-B} \prod_{i=1}^{D} X_i }.
\ee
The linearization around $X=1$ yields
\be
\label{beta-crit-qu}
\beta_c^{\mathrm{qu}}= \atanh\left(1/\lambda\right).
\ee

Surprisingly, the quenched critical value coincides with
the one that is obtained by solving the annealed Ising
model on the  {\em combinatorial Erd\H{o}s-R\'enyi}  random graph,
denoted  $\text{cER}(\lambda n/2)$,
with $n$ vertices and a {\em fixed} number of edges $\lambda n/2$
placed uniformly at random. 
The annealed magnetization of the combinatorial{ Erd\H{o}s-R\'enyi}  random graph
satisfies yet another mean-field equation (see Supplementary Material)
{\be\label{eq-magn-cER}
M^{\text{cER}} = \tanh\Big[ \frac{\lambda(1-\e^{-2\beta})M^{\text{cER}}}{2+(1-\e^{-2\beta})((M^{\text{cER}})^2-1)} + B \Big]\,.
\ee
}
The linearization around zero gives
\be
\beta_c^{\text{cER}}= \beta_c^{\mathrm{qu}}.
\ee
We observe that, although the bER
and the cER
are asymptotically equivalent random graph
models (in particular they both have 
Poisson degrees),
their annealed magnetization satisfies 
different equations yielding different critical
temperatures. 
We show in the Supplementary Material that this difference arises from 
the fact that in the cER, the number
 of edges is fixed, whereas annealing macroscopically increases the number of edges in the bER.
%{\color{red} [Say more???]}
  
\bigskip 
\noindent
{\bf Model with inhomogeneities.} 
We now go to a more general setting that allows
to treat inhomogeneities described by general 
degree distributions (beyond the Poissonian
case).

 We first consider the
 {\em configuration model with fixed degrees}, denoted by $\mathrm{CM(d)}$ that is 
 obtained by prescribing the degree values $\mathrm{d}= (d_i)_{i\in \chr{[n]}{}}$
 and connecting the vertices uniformly at random
 \cite{remco}.
 In the Supplementary Material we show that,
 denoting by $D$ the degree of a uniformly chosen
 vertex,  the annealed magnetization is
 \be
\label{configuration-magn-y}
M^{\mathrm{CM(d)}}(\beta,B) = \lla \tanh\Big(\beta y D + B\Big)\rra,
\ee
where
$y \in (0,1)$  is a solution to 
\be
\label{configuration-fixed-y}
\frac{1-\e^{-4\beta y}}{1+\e^{-4 \beta y} - 2 \e^{-2\beta(1+y)}} = \lla D \tanh\Big(\beta y D + B\Big)\rra / \lla D\rra.
\ee
Comparing \eqref{configuration-fixed-y} and \eqref{dor-fixed} we see once more that the
``annealed network approach'' correctly predict a mean-field behaviour for the annealed
magnetization, \chr{but}{} that the mean-field equation
for $y$ is again \chr{quite}{} different.
From the linearization of equation \eqref{configuration-fixed-y} around $y=0$ 
one finds that the annealed critical point $\beta_c^{\mathrm{CM(d)}}$ of the configuration model with
prescribed Poissonian degrees is
\be
\label{crit-value-CMd}
\beta_c^{\mathrm{CM(d)}}= \beta_c^{\mathrm{qu}},
\ee
which is consistent with the claim that fixing the number of edges
recovers the quenched critical temperature.

If instead the configuration model is constructed by
considering random i.i.d.\  degrees $D_i$
(denoted by $\mathrm{CM(D)}$), then the situation
drastically changes. The additional randomness of the degrees
implies that only degrees distributions with
exponential tails are possible. 
Indeed, by considering the configuration $\sigma$
with all spins up, one immediately obtains
the bound
\be
\label{diciotto}
 \lla \e^{\beta \sum_{i\in V} D_i/2}\rra \le \e^{-n|B|} \lla Z_n\rra \le 2^n  \lla \e^{\beta \sum_{i\in V} D_i/2}\rra\,.
\ee
The annealed free energy is thus \chr{only}{} well-defined in the thermodynamic limit 
if  $\lla \e^{\beta D/2} \rra < \infty$.
Assuming this to be the case, then  as shown in the Supplementary Material,  
the annealed magnetization reads
 \be
\label{configuration-magn-y-iid}
M^{\mathrm{CM(D)}}(\beta,B) = \lla \tanh\Big(\beta y D_{\beta} + B\Big)\rra,
\ee
where
$y \in (0,1)$  is a solution to 
\be
\label{configuration-fixed-y-iid}
\frac{1-\e^{-4\beta y}}{1+\e^{-4 \beta y} - 2 \e^{-2\beta(1+y)}} = \lla D_{\beta} \tanh\Big(\beta y D_{\beta} + B\Big)\rra/\lla D_{\beta}\rra.
\ee
Here $D_{\beta}$ is the new law that arises from the law of $D$ as a consequence of 
the randomness of the degrees. Indeed, in the presence of i.i.d. degrees that 
are copies of a random variable $D$ with distribution $\p=(p_k)_{k \ge 1}$, i.e. $\pr (D=k)=p_k$, 
the annealed `pressure' ($= -f/\beta$ with $f$ the annealed free energy) is   \cite{C}
    \eqn{\label{pressCMD}
\varphi^{\mathrm{CM(D)}}(\beta,B)= \sup_{{\q}} \left [ \varphi^{\mathrm{CM(d)}}(\beta,B; \q)- H(\q | \p) \right],
}
where $\varphi^{\mathrm{CM(d)}}(\beta,B; \q)$ denotes the pressure of the configuration model  with  a {\em deterministic} degree distribution $\q$, and 
$H(\q | \p)$ is the relative entropy of $\q$ with respect to $\p$
\be
H(\q | \p) = \sum_k {q_k}\log{\frac{q_k}{p_k}}.
\ee
The equations \eqref{configuration-magn-y-iid}, \eqref{configuration-fixed-y-iid} are then
obtained by deriving w.r.t. the external field $B$. 
To identify the critical temperature, one takes $B\searrow 0$, 
in which case the law of $D_{\beta}$ turns out to be 
a $\beta$-dependent exponential tilting of the degree-distribution $D$,
\be\label{qtilting}
q_k(\beta)=p_k \cosh(\beta)^{k/2}/c(\beta),
\ee 
with $c(\beta)=\lla \cosh(\beta)^{D/2}\rra.$
Thus, \chr{since}{} $\cosh(\beta)>1$, under the annealed measure 
of the configuration model, the typical graph in the case
of random i.i.d.\ degrees re-arranges itself (compared to the
case of deterministic degrees) in order to 
maximize the ferromagnetic alignment of spins, \chr{and it does so}{} by increasing the number of edges.

In particular, when the degree $D$ is Poissonian with mean $\lambda$,
the tilted degree $D_{\beta}$ is \chr{again a}{} a Poisson random variable with mean $\lambda\sqrt{\cosh(\beta)}$. 
The linearization of \eqref{configuration-fixed-y-iid} around $y=0$ then yields
an implicit equation for the critical inverse temperature
\be
\beta_c= \atanh\Big(\frac{1}{\lambda\sqrt{\cosh(\beta_c)}} \Big),
\ee
whose solution $\beta_c^{\mathrm{CM(D)}} $ is 
\be
- \log (2 \lambda^2) + \log   \left[1+\sqrt{1+4 \lambda^4}+\sqrt{2 +2 \sqrt{1+4 \lambda^4}}\right].
\ee
Comparing to \eqref{crit-value-CMd}, we see that while the annealed $\mathrm{CM(d)}$
with fixed Poissonian degrees has a phase transition only when a giant connected component 
exists $(\lambda>1)$, the $\mathrm{CM(D)}$ with random Poissonian
degrees has a finite critical temperature for {\em all} $\lambda >0$.

\chr{In Fig.\ 1, we collect}{} the results obtained so far.  
For all the random networks with {\em Poisson degree distribution},
the quenched critical temperature  
is given by \eqref{beta-crit-qu}, but we have 4 different values of the annealed 
critical temperatures. 
In particular, the combinatorial Erd\H{o}s-R\'enyi (cER) and the configuration model $\mathrm{CM(d)}$,  both having  a fixed number of edges, reproduce the quenched critical value, whereas the binomial Erd\H{o}s-R\'enyi (bER) and the configuration model $\mathrm{CM(D)}$ with a fluctuating number of edges, have a critical value that is model-dependent (and different from that \chr{of the}{} mean-field ``annealed network approach''). 
\bigskip
%\begin{table}[htp]
%\begin{center}
%\begin{tabular}{|c|c|}
%\hline
%Model & $\beta_c(\lambda)$\cr
%\hline\hline
%A & $\asinh (\lambda^{-1})$ \cr
%\hline
%B & $(\lambda+1)^{-1}$ \cr
%\hline
%C, D,  E & $\atanh(\lambda^{-1})$ \cr
%%\hline
%%D & $\atanh(\lambda^{-1})$ \cr
%%\hline
%%E  & $\atanh(\lambda^{-1})$ \cr
%\hline
%F & $-\log (2 \lambda^2) + \log   \left[1+f(\lambda)+\sqrt{2 +2 f(\lambda)}\right]$ \\
%\hline 
%\end{tabular}
%\end{center}
%\caption{\label{tab1}  Annealed critical points  $\beta_c(\lambda)$ for the models considered above.  
%Model A: Gilbert random graph with independent edges with probability $\lambda/n$.
%Model B: the ``annealed network approach'', defined by the Hamiltonian \eqref{H_Dor-xxx}.
%Model C: quenched model.
%Model D: Erd\H{o}s-R\'enyi  random graph, with  $\lambda n/2$ edges placed uniformly at random.
%Model E: Configuration model with prescribed Poisson degrees. 
%Model F: Configuration Model  with i.i.d. Poisson degrees (here $f(\lambda)= \sqrt{1+4 \lambda^4}$).}
%\end{table}
%

\begin{figure}[ht!]
\label{Fig1}
\begin{center}
~\vskip-0.5cm
\includegraphics[width=0.4\textwidth]{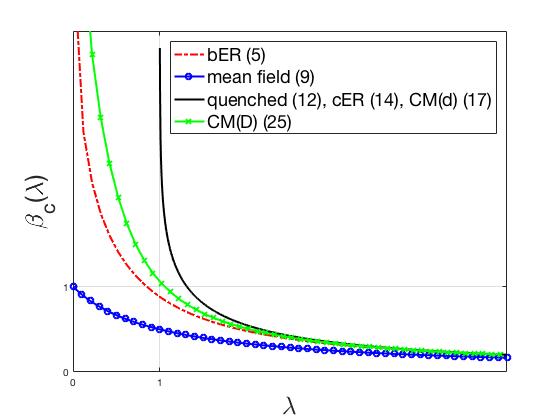}
\end{center}
\vskip -0.5cm
\caption{Annealed critical points $\beta_c(\lambda)$ for  models with degree distribution Poisson($\lambda$).
Points are guides to the eye.  
%In Model F $g(\lambda)= 1 + \sqrt{1+4 \lambda^4}+\sqrt{2 + 2 \sqrt{1+4 \lambda^4}}$
}
\end{figure}

\noindent
{\bf Presence or absence of power-law universality class.} We now analyze the annealed critical exponents.
One immediately sees that for homogeneous networks  (i.e., Poissonian degree distribution), 
the critical exponents are those of \chr{the}{} Curie-Weiss model. Therefore, we concentrate on 
inhomogeneous networks and, for the sake of space, we consider the configuration model.

We start from the case of {\em fixed degrees}: by Taylor expansion of Eq.\ \eqref{configuration-magn-y}, 
\eqref{configuration-fixed-y}, \chr{we now}{} obtain a critical temperature 
$\beta_c^{\mathrm{CM(d)}}= \atanh\left(\frac{\langle D\rangle}{\langle D(D-1)\rangle}\right)$. Thus the annealed system
has a ferromagnetic phase transition when $\langle D^2\rangle<\infty$ and is always in the
ferromagnetic phase when $\langle D^2\rangle=\infty$. As for the critical exponents, \chr{we find}{} 
those of the mean-field universality class, provided that 
$\langle D^4\rangle<\infty$. If this condition is not met,
then new universality classes arise \cite{leone, doro-zero, DGH1}. 
For instance, for power-law distributed degrees, 
i.e.\ $p_k \sim k^{-\tau}$ with an exponent $3 < \tau < 5$, \chr{we find}{}
$$
\alpha= \frac{5-\tau}{\tau-3},\qquad \beta= \frac{1}{\tau-3}, \qquad \gamma = 1, \qquad \delta = \tau-2.
$$
This scenario of a family of universality classes (labeled by \chr{the degree power-law}{} exponent $\tau$) coincides with 
what was found for all quenched networks with a locally-tree like structure \cite{leone, doro-zero, DGH1}.

We now move to the configuration model with {\em random i.i.d.\ degrees}.
Taylor expansion of \eqref{configuration-magn-y-iid} and \eqref{configuration-fixed-y-iid}
identifies the critical inverse temperature $\beta_c^{\mathrm{CM(D)}}$ as the solution of the equation
$$
\beta= \atanh \left ( \frac{\lla D_{\beta}\rra}{\lla D_{\beta}(D_{\beta}-1)\rra}\right) .
$$
As we \chr{have}{} already remarked, for power-law degrees the free energy simply blows up. 
Thus, we have to restrict to degree distributions with exponential tails, in
which case, the free energy diverges when $\beta$ is large, but not when it is small.
In this case, the critical value $\beta_c^{\mathrm{CM(D)}}$ 
is strictly smaller than the value $\beta$ where the free energy explodes.
Then, provided that $\lla \e^{\beta_c^{\mathrm{CM(D)}} D/2}\rra < \infty$
(cf. \eqref{diciotto}), 
 the tilted degree distribution $q(\beta)$ in \eqref{qtilting} {\em always}
has exponential tails, since $\cosh(\beta) < \e^{\beta}$.
Therefore, the empirical degree 
distribution $q(\beta, B)$ \chr{of the random graph under the annealed Ising model}{} with a non-zero field $B$, 
close to the critical point, has exponential tails. 
As a result, power-law degree distributions cannot occur,
and thus the critical exponents are all equal to those of the Curie-Weiss model.
In this case, there exists {\em only one universality class,} compared to the several ones for
the setting of deterministic degrees. 

{\bf Acknowledgements.} We thank J.F.F.\ Mendes for useful discussions.
RvdH acknowledges financial support from Gravitation-grant NETWORKS-024.002.003.

\newpage

{\bf SUPPLEMENTARY MATERIAL}

\section{Annealed combinatorial Erd\H{o}s-R\'enyi}
%The quenched critical point for the Erd\H{o}s-R\'enyi random graph $\text{bER}(\lambda/ n)$ of size $n$ and with edge probability $\lambda/n$ equals $\beta_c^{\rm \sss qu}=\atanh(1/\lambda)$. Further, the annealed critical value equals $\beta_c^{\rm \sss an}=\asinh(1/\lambda)$. The fact that these are different again confirms the belief that the critical value is determined by the number of edges. 
Here, we investigate the annealed Ising model on the Erd\H{o}s-R\'enyi random graph $\text{cER}(\lambda n/2)$ of size $n$ with a fixed number $m= \lambda n/2$ of edges placed uniformly at random, for which we prove that the critical value equals the quenched critical value.

Let us denote $Z^{\text{cER}}_{n,+}(k)$  the partition function where we fix $|\sigma_+|=k$ with $\sigma_+$  the subset of sites with positive spin. Then,
recaling the Hamiltonian \eqref{ham}, the annealed partition function equals
	\eqn{
	\lla Z^{\text{cER}}_n\rra = 
	\lla\sum_{\sigma}\e^{H^{\text{cER}}(\sigma)}\rra
	=\sum_{k=0}^n  \lla Z^{\text{cER}}_{n,+}(k)\rra.
	}
Using 
$$
H^{\text{cER}}(\sigma) = m - 2 e(\sigma_+,\sigma_-) + B (2|\sigma_+| -n)
$$ 
where $e(\sigma_+,\sigma_-)$ is the number of edges connecting $\sigma_+$ to $\sigma_-$, we get
\eqn{
	\lla Z^{\text{cER}}_{n,+}(k)\rra
%	= \lla \sum_{\sigma \atop |\sigma_+|= k } e^{\beta m - 2 \beta e(\sigma_+,\sigma_-) + \beta B n (2k-1)}\rra.
		= \lla \sum_{|\sigma_+|= k } \e^{\beta m - 2 \beta e(\sigma_+,\sigma_-) + \beta B n (2k-1)}\rra.
	}
By adding edges one-by-one, we see that, with $m=\lambda n/2$ and $N=n(n-1)/2$, 
we get
	\eqn{
	\lla Z^{\text{cER}}_{n,+}(k)\rra
	= {n\choose k} \Big(1+(\e^{-2\beta}-1)\frac{k(n-k)}{N}\Big)^m.
	}
Here we  ignored possible double additions of edges, which is not relevant in the thermodynamic limit.
Therefore, the annealed `pressure'  $\varphi^{\text{cER}}  = \lim_{n\to \infty}\frac{1}{n} \log \lla Z_n^{\text{cER}} \rra$ equals
	\eqn{
	\label{ann-fixed-press}
	{\varphi}^{\text{cER}}_{\beta,B}(\lambda)=\sup_{t\in [0,1]} \Big[ I(t)+\hat{\varphi}_{\beta,B}(t)\Big]
	}
where 
	\eqn{\label{phihat}
	I(t)=-t\log{t}-(1-t)\log(1-t)	
	}
	\eqan{
	\label{phihat2}
	\hat{\varphi}_{\beta,B}(t)&= (\lambda/2)\Big[\beta+\log\Big(1+(\e^{-2\beta}-1)2t(1-t)\Big)\Big]\nn\\
	&\qquad+B(2t-1).
	}

Optimizing over $t$ in \eqref{ann-fixed-press} yields that the optimizer $t^\star$ is the solution to

		\eqn{\label{eq-stat-cER}
	\log \frac{(1-t)}{t}+2B+\frac{\lambda(\e^{-2\beta}-1)(1-2t)}{1+(\e^{-2\beta}-1)2t(1-t)}=0.
	}		
	Calling $P_{\beta,B}(t) = I(t) + \hat{\varphi}_{\beta,B}(t)$, 
	the magnetization is $M(\beta, B)= \frac{\partial }{\partial B} P_{\beta, B}(t^\star(\beta,B)) $, then
	$$
	M(\beta, B)  =2 t^\star -1.
	$$
Equation \eqref{eq-magn-cER} for the magnetization of the combinatorial Erd\H{o}s-R\'enyi model can be obtained by substituting  $t^\star=(M+1)/2$ into 
 	\eqref{eq-stat-cER}. We get
	\eqn{
	\log \left ( \frac{1+M}{1-M}\right ) = 2 B + \frac{2 \lambda (\e^{-2\beta}-1) M  }{ 2+  (\e^{-2\beta}-1) (M^2 -1) }
	}
	which is \eqref{eq-magn-cER}.

The critical value $\beta_c$ \chr{satisfies}{}
	\eqn{
	\frac{\partial^2}{\partial t^2} P_{\beta_c,0^+}(t)\big|_{t=\tfrac{1}{2}}=0.
	}
Computing the second derivative gives
	\eqan{
	\frac{\partial^2}{\partial t^2} P_{\beta, 0^+}(t)
	&=-\frac{1}{t}-\frac{1}{(1-t)}
	-\frac{2\lambda(\e^{-2\beta}-1)}{1+(\e^{-2\beta}-1)2t(1-t)}\nn\\
	&\quad-\frac{\lambda}{2}\frac{\Big((\e^{-2\beta}-1)2(1-2t) \Big)^2}{(1+(\e^{-2\beta}-1)2t(1-t))^2}.
	}	
This derivative computed at $t=\tfrac{1}{2}$ gives
	$$
	-2-\frac{\lambda(\e^{-2\beta}-1)}{1+(\e^{-2\beta}-1)/2}=0,
	$$
or 
	$$
	\lambda \tanh(\beta)=1.
	$$
	Therefore, the critical value $\beta_c^{\text{cER}}$ of the combinatorial Erd\H{o}s-R\'enyi random graph with $n$ vertices and  $\lambda n/2$ edges  equals $\atanh(1/\lambda)$.

	\section{Annealed binomial\\ Erd\H{o}s-R\'enyi}
	
	The annealed  binomial Erd\H{o}s-R\'enyi random graph (with fluctuating number of edges) was solved in \cite{GGHP2, DGGHP}
	by a direct mapping to the inhomogenous Curie-Weiss model.
	Here we show that the solution arises from the combinatorial Erd\H{o}s-R\'enyi (cER) random graph (with fixed number of edges) 
	via the total probability formula.
	We have
\[
\lla Z^{\text{bER}}_n \rra = \sum_{k\geq 1} \lla Z_n \rra_{\text{cER}(k)} \pr(\#\{\textrm{edges in } \text{bER}(\lambda/n)\}=k)
\]
where $\lla \cdot \rra_{\text{cER}(k)}$ is the expectation w.r.t.\ the cER random graph with fixed number of edges $k$. 
Now,
\eqn
{\lla Z_n\rra_{\text{cER}(n\mu/2)}  = \exp({n[\varphi^{\text{cER}}_{\beta,B}(\mu)+o(1)]}),
}
and
\eqn{
\pr \left (\#\{\textrm{edges in } \text{bER}(\lambda/n)\}=n \mu/2 \right) = \e^{-n[S_{\lambda}(\mu)+o(1)]},
}
where 
$S_{\lambda}(\mu)$ is the relative entropy of the Binomial distribution $\text{Bin}(N,\mu/n)$
with respect to Binomial distribution $\text{Bin}(N,\lambda/n)$ \chr{given by}{}
\[
S_{\lambda}(\mu) = \frac{1}{2}  \left( \mu\log \frac{\mu}{\lambda}+\mu-\lambda\right).
\]
Considering the pressure $\varphi^{\text{bER}}  = \lim_{n\to \infty}\frac{1}{n} \log \lla Z_n^{\text{bER}} \rra$,
%\eqn{
%\phi_{\beta,B} (\lambda) = \lim_{n\rightarrow \infty}\frac{\log \expec[Z_n]}{n} \,,
%}
a saddle point argument (\chr{or}{} Varadhan's lemma)
implies
\eqn{\label{optimERbin}
\varphi^{\text{bER}}_{\beta,B} (\lambda) = \sup_{\mu>0} \Big\{ 	{\varphi}^{\text{cER}}_{\beta,B}(\mu) - S_{\lambda}(\mu) \Big\},
}
with
\eqn{
{\varphi}^{\text{cER}}_{\beta,B}(\mu) = P_{\beta,B}(t^{\star}(\mu)),
}
where the optimizer $t^\star = t^\star (\mu)$ satisfies (cf.\ \eqref{eq-stat-cER})
\eqn{\label{station}
	\log\frac{t^\star}{1-t^\star} -2B = \frac{\mu(\e^{-2\beta}-1)(1-2t^\star)}{1+(\e^{-2\beta}-1)2t^\star(1-t^\star)}.
	}	
The stationarity condition for  \eqref{optimERbin}, i.e., 
\eqn{\label{statlambda}
\frac{\partial \varphi^{\text{cER}}_{\beta,B}}{\partial \mu} (\mu)\equiv \frac{\partial P_{\beta,B}}{\partial \mu}(t^\star (\mu))= \frac{\partial S_\lambda}{\partial \mu}(\mu)
}
yields the implicit equation for the optimizer ${\mu}^\star={\mu}^\star (\lambda, B)$ \chr{as}{}
\eqn{
 \beta + \log \left (1+ (\e ^{-2 \beta}-1) 2 t^{\star} (1-t^{\star}) \right ) = \log \frac{\mu}{\lambda}},
from which we get
$$
\mu^{\star}= \lambda \e^\beta [1+(\e^{-2 \beta}-1) 2 t^{\star}(1-t^{\star}) ].
$$
Substituting $\mu^{\star}$ in \eqref{station}, we get
\eqn{\nonumber
\log \frac{t^\star }{1-t^\star} = \lambda (\e^{- \beta}-\e^{\beta})(1-2t^\star) +2B.
}
Since the magnetization is $M^{\text{bER}}=2t^\star-1$ (as we show below), we rewrite the previous equation as
\[
\atanh(M^{\text{bER}}) = \lambda \sinh(\beta) M^{\text{bER}} +B,
\]
which is 
%or equivalently
%\eqn{
%m=\tanh(\lambda \sinh(\beta) m + B).
%}
 the equation \eqref{magn-er-bin} for the magnetization of the bER. In order to
 check that $M^{\text{bER}}=2t^\star-1$, we write $\varphi^{\text{bER}}_{\beta, B}(\lambda)={\varphi}^{\text{cER}}_{\beta,B}({\mu}^\star(\lambda, B)) - S_{\lambda}({\mu}^\star(\lambda, B))$ 
 and compute $M^{\text{bER}}=\frac{\partial \varphi^{\text{bER}}_{\beta, B}}{\partial B}$. We have
 \eqan{\label{derphil}
 \frac{
 \partial \varphi^{\text{bER}}_{\beta, B}}{\partial B}(\lambda)&=  \frac{\partial \varphi^{\text{cER}}_{\beta, B}}{\partial B}({\mu}^\star(\lambda, B)) + \left [ \frac{\partial \varphi^{\text{cER}}_{\beta, B}}{\partial \mu}({\mu}^\star(\lambda, B))\right. \nn \\
 & \left. - \frac{\partial S_\lambda}{\partial \mu}({\mu}^\star(\lambda, B)) \right ] \frac{{\mu}^\star(\lambda, B)}{ \partial B} .
 }
The term in square brackets  vanishes because ${\mu}^\star(\lambda, B)$  satisfies \eqref{statlambda}. On the other hand, the partial derivative with respect to $B$ of $\varphi^{\text{cER}}_{\beta, B}$ is
\eqan{
& \frac{\partial \varphi^{\text{cER}}_{\beta,B}}{\partial B}(\mu) = \frac{\partial P_{\beta, B, \mu} }{\partial B}(t^\star(\mu,B))+ \frac{\partial P_{\beta, B, \mu} }{\partial t}(t^\star(\mu,B))\nn \\
& \times \frac{\partial t^\star(\lambda, B)}{\partial B} = 2  t^\star(\lambda, B) - 1,
}
since the derivative of $P_{\beta, B, \mu}$ w.r.t.\ $t$ vanishes at the optimizer $ t^\star(\lambda, B)$ and $\frac{\partial P_{\beta, B, \mu} (t)}{\partial B}\equiv \frac{\partial \hat{\varphi}_{\beta,B}(t) }{\partial B }= 2t-1$,
see \eqref{phihat2}.

%Observe that 
%\eqan{
%\partial_d J_\lambda &= \frac{1}{2}(\log \lambda -\log d) \nn \\
%\partial_d  {\varphi}_{\beta,B} &= \frac{1}{2} \log\Big(\e^{\beta}+(\e^{-\beta}-\e^{\beta})2t_*(1-t_*)\Big), \nn
%}
%where $t_*$ is the solution of $I'(t)+ \hat{\varphi}_{\beta,B}'(t)=0$, i.e.
%\eqn{
%\log \frac{t}{1-t} = d \frac{(\e^{- \beta}-\e^{\beta})(1-2t)}{\e^{\beta}+(\e^{- \beta}-\e^{\beta})2t(1-t)} +2B.
%}
%The optimizer $d_*$ of \eqref{eod} is the solution of $\partial_d(J_{\lambda}+{\varphi}_{\beta,B})=0$, and thus
%\eqn{
%d_*=\lambda \Big(\e^{\beta}+(\e^{-\beta}-\e^{\beta})2t_*(1-t_*)\Big).
%}
%Combining the last two equations, 
%\eqn{
%\log \frac{t_*}{1-t_*} = \lambda (\e^{- \beta}-\e^{\beta})(1-2t_*) +2B.
%}
%Moreover, the magnetization $m=2t_*-1$, and so 
%\[
%\atanh(m) = \lambda \sinh(\beta) m +B,
%\]
%or equivalently
%\eqn{
%m=\tanh(\lambda \sinh(\beta) m + B).
%}
%We finally derive the equation \eqref{magn-er-bin}.
%
%\newpage
%\phantom{x}
{
    \section{Annealed configuration model with deterministic Degrees}
In \cite{C} we have shown that the pressure of the annealed configuration model with deterministic degrees is
\eqn{\label{pressCMd}
\varphi^{\mathrm{CM(d)}}(\beta,B)= \frac{\beta \lla D\rra}{2}+  G((s^\star_k)_{k\ge 1},B),
}
where $D$ is the  degree distribution, and $G$ is a function of the infinite-dimensional vector $(s_k)_{k\geq 1}\in (0,1)^\N$ given by
	\eqan{
	\label{G(sk)-def}
	G((s_k)_{k\geq 1}, B)&=\sum_{k} p_k I(s_k) + B \Big(2\sum_{k} s_kp_k -1\Big) \nn\\
	& + \lla D \rra  F_{\beta}\Big(\frac{\sum_{k } k p_k s_k}{ \lla D\rra }\Big).
	}
Here  $p_k=\pr(D=k)$ and $F_\beta$ is a function that we do not need to make explicit here  (see \cite{C}). The \chr{vector of optimizers}{} $(s^\star_k)_{k\ge 1}$ in \eqref{pressCMd} is defined as
\eqn{\label{sstar} 
s^\star_k(B)=(w^k \e^{-2 B}+1)^{-1},
}
 where, for $B>0$, $w=w(\beta,B)$ is a solution in $(\e^{-2\beta},1)$ to
	\eqn{
	\label{v-def}
	\frac{1 - \e^{-2 \beta}w }{1+w^2-2\e^{-2 \beta} w } =  \lla \left(1+w^{D^\star} \e^{-2B} \right)^{-1}\rra,
    }
and $D^\star$ is the size-biased  random variable given by $\pr (D^\star=k)=kp_k/ \lla D\rra$. Thus,  the magnetization $M^{\mathrm{CM(d)}}$ can be computed as
    \eqan{\label{man}
    M^{\mathrm{CM(d)}} &= \frac{d}{dB} G((s^\star_k(B))_{k\geq 1}, B)  =   \frac{\partial G}{\partial B }((s^\star_k(B))_{k\geq 1}, B) \nn  \\
    & + \sum_{k\geq 1} \frac{\partial G}{\partial s_k }((s^\star_k(B))_{k\geq 1}, B)\, \frac{d s^\star_k(B) }{d B}\nn \\
    & = 2 \sum_{k } s^\star_k(B)p_k -1=  \sum_{k } \frac{\e^{2 \beta} - w^k}{ \e^{2 \beta} + w^k}p_k,
    }
where we use \eqref{sstar} and the fact that the partial derivatives $\frac{\partial G}{\partial s_k}$ vanish at $(s^\star_k)_{k\ge 1}$, (see  \cite{C}). 
Since $\tanh (x+y) = \frac{\e^{2x}-\e^{-2y}}{\e^{2x}+\e^{-2y}}$, defining $y$ by $w=\e^{-2 \beta y}$, we can rewrite \eqref{man} as
    $$
    M^{\mathrm{CM(d)}} = \lla  \tanh (  \beta y  D+ B) \rra,
    $$
which is \eqref{configuration-magn-y}. In the same fashion, writing $(1+w^{D^\star} \e^{-2B} )^{-1}$ as $\frac{1}{2} \tanh (\beta y D^\star +B) + \frac 1 2$
    in \eqref{v-def} we obtain 
    $$
    \frac{1 - w^2 }{1+w^2-2\e^{-2 \beta} w } = \lla \tanh \big (\beta y D^\star+B \Big )\rra,
    $$
\chr{which, in turn,}{} can be transformed into \eqref{configuration-fixed-y} by substituting $w=\e^{-2 \beta y}$ and using that $D^\star$ is the size-biased degree. This proves our statements concerning the magnetization of the configuration model with fixed degrees $\text{CM(d)}$. 
    }

    \section{Annealed configuration model with random degrees}

In the case in which the degrees are  i.i.d.\ copies of a random variable $D$ with distribution $\p=(p_k)_{k \ge 1}$, i.e., $\pr (D=k)=p_k$, the annealed pressure is  
    \eqn{\label{pressCMD2}
\varphi^{\mathrm{CM(D)}}(\beta,B)= \sup_{{\q}} \left [ \varphi^{\mathrm{CM(d)}}(\beta,B; \q)- H(\q | \p) \right],
}
where $\varphi^{\mathrm{CM(d)}}(\beta,B; \q)$ denotes the pressure of the configuration model  with {deterministic}  degree distribution $\q$ and 
$H(\q | \p)$ is the relative entropy of $\q$ with respect to $\p$. The variational representation of the pressure \eqref{pressCMD2} can be rewritten as 
\be\label{supwq}
\varphi^{\mathrm{CM(D)}}(\beta,B)= \sup_{w,{\q}} R_{\beta, B}(w,\q),
\ee
with (cf.\ \eqref{pressCMd})
\eqan{
R_{\beta, B}(w,\q) &=-H(\q | \p) + \frac{\beta \lla D(\q)\rra}{2} \nn \\
& +  G((s_k(w,B))_{k\ge 1},B;\q),\nn
}
where
 $s_k(w,B)=\e^{2B}/(\e^{2B}+w^k)$ and $G((s_k)_{k\geq 1}, B; \q)$ is defined as in \eqref{G(sk)-def} with $\p$
replaced by $\q$ and $D$ by $D(\q)$.  The latter is the degree random variable with distribution $\q$.
Denoting by $(w^\star, \q^\star)$ the optimizer of the variational problem \eqref{supwq}, we write
    \eqan{\label{pressCMDS}
\varphi^{\mathrm{CM(D)}}(\beta,B)&= R_{\beta, B}(w^\star,\q^\star)= -H(\q^\star | \p) + \frac{\beta \lla D(\q^\star)\rra}{2} \nn \\
& +  G((s_k(w^\star,B))_{k\ge 1},B;\q^\star).
}
The relation between $\p$ an  the optimizing distribution $\q^\star$ can be obtained by a stationarity condition which, $\q$ being 
a probability \ch{mass function,}{} is given by
\be\label{lagrange}
\frac{\partial R_{\beta, B}(w^\star,\q^\star) }{\partial q_k}=\zeta,
\ee
for some Lagrange multiplier $\zeta$. Computing the derivatives, we obtain
\eqan{\label{qstarp}
\log(q_k^\star/p_k) &= k \left [ \frac \beta 2 + F_\beta(\stq) + F^\prime_\beta(\stq)(s^\star_k-\stq) \right]\nonumber \\
& + I(s^\star_k)+ 2 s^\star_k B + \zeta,
}
where $\stq=(\sum_k k s^\star_k q^\star_k)/\lla D(\q^\star)\rra$ is the average of the vector $s_k^\star=s_k(w^\star,B)$ w.r.t.\
the size-biased distribution of $\q^\star$. 
The stationarity condition for $(s^\star_k)_{k\ge 1}$, i.e. $\frac{\partial G((s^\star_k)_{k\ge 1},B;\q^\star)}{\partial s_k}=0$ is
$$
q^\star_k \cdot \large ( I^\prime(s^\star_k) + 2 B + k F^\prime_\beta (\stq)\large) =0.
$$
From this equation we get  $F^\prime_\beta (\stq)$ that inserted in  \eqref{qstarp} yields
\eqan{\label{qstarpbis}
\log(q_k^\star/p_k) &= k \left [ \frac \beta 2 + F_\beta(\stq) + I(s^\star_k)+ I^\prime(s^\star_k)(s^\star_k-\stq) \right]\nonumber \\
& + 2 \stq B + \zeta.
}
The implicit relation \eqref{qstarp}  (or \eqref{qstarpbis}) can be made explicit for vanishing field $B\searrow 0$. 
In this case,  since $(s^\star_k)_{k \ge 1} \to (\tfrac1 2)_{k\ge 1}$ and $F_\beta(\tfrac{1}{2})=-\beta/2 + \frac 1 2 \log \cosh(\beta)$,
equation  \eqref{qstarp} yields
$$
\log(q_k^\star/p_k) - \frac k 2  \log \cosh(\beta) =\zeta,
$$
which is \eqref{qtilting}.\\
{
In order to show \eqref{configuration-magn-y-iid} and \eqref{configuration-fixed-y-iid}, we start from \eqref{pressCMDS} and, observing that $\frac{\partial  R_{\beta, B} }{\partial B}=\frac{\partial  G((s_k)_{k\geq 1}, B; \q) }{\partial B}$, compute
\eqan{
 &M^{\mathrm{CM(d)}} = \frac{d}{dB} R_{\beta, B}(w^\star(B),\q^\star(B))\nonumber \\
 &= \sum_ k \frac{\partial R_{\beta, B}(w^\star,\q^\star) }{\partial q_k} \frac{\partial q_k^\star}{\partial B} + \frac{\partial  G((s_k^\star)_{k\ge 1},B;\q^\star) }{\partial B}\nonumber \\
 &  = 2 \sum_{k\ge 1} s^\star_k q^\star_k -1
}
where we use \eqref{lagrange} and the fact that $\sum_k  \frac{ \partial q^\star_k(B)}{\partial B}= \frac{ \partial}{\partial B} \sum_k q^\star_k(B)  =  \frac{ \partial }{\partial B} 1=0$. From this point on, the proof 
proceeds as in the case of fixed degrees (see \eqref{man}), with $D$ replaced by $D_\beta \equiv D(\q^\star)$.
}

\end{document}